# Higher education funding: The value of choice

Limor Hatsor and Ronen Bar-El

E-mail: limor.hatsor@gmail.com

**Abstract**

An alternative to the dependence on traditional student loans may offer a viable relief from the tremendous burden that those loans usually incur. This article establishes that it is desirable for governmental intervention to grant students 'more choice' in their funding decisions by allowing them to have portfolios, mixtures of different types of loans. To emphasize this point, a model is presented of a situation where students invest in higher education while facing uncertainty about their individual earning potential. The model reveals that when students are allowed to have portfolios of loans, some of them indeed take the opportunity and diversify their loans, benefiting themselves, but also improving the loan terms of other students. Therefore, when governments organize student loans, they should consider providing students with more choice in their funding decisions.

I am thankful to participants at the workshops at Tel-Aviv University, Bar Ilan University, Ben-Gurion University, and participants at PET13 conference as well as Konstantin Malsev, Miriam Hatzor, Rachel Herchy Linnewiel, Revital Huber Shalem and Glynn Mathis. I also thank the Sanbar Institute for Public Economics for funding this project.

# 1. Introduction

This article introduces a well-known tool of financial markets into student loans. Allowing students to choose their own portfolio, or mixture, of different types of loans is suggested as a remedy for the drawbacks of traditional funding channels.

The issue of funding higher education is an ongoing worldwide debate. On the one hand, abundant empirical evidence has highlighted its important role for individuals and countries (Barro, 1998; Bassanini and Scarpenta, 2001; Restuccia and Urrutia, 2004). Correspondingly, investment of governments in education has increased substantially in members of the 'Organization for Economic Co-operation and Development' (OECD) during the second half of the twentieth century (Checchi, 2006; Greenaway and Haynes, 2003). On the other hand, budgetary pressures and other considerations have recently led many countries to shift from public higher education funding (through various forms of income support transfers) to private funding (based on student loans).

Thus, decisions about how to finance higher education (referred to as 'funding decisions' hereinafter) have become an individual choice in terms of student loans. This begs two questions, which are the focus of this study: 1) How do individuals react to being granted 'more choice' by allowing the use of portfolios as an addition to traditional loans? 2) Is 'more choice' beneficial for the economy? These are practical questions for governments that consider different policies of higher education funding.

The model presented later in this paper establishes that 'more choice' is valuable for certain students who prefer to mix loans, and at the same time beneficial for other students also. Even students who prefer using only one type of a loan may benefit. While they do not use the opportunity to diversify loans, their borrowing terms hinge on the funding decisions of other students. Specifically, their borrowing terms may improve when other students decide to mix loans. Therefore, it is important that governments not only guarantee access for students to loans (see a literature review on students' borrowing constraints in Eckwert and Zilcha, 2015) but also ensure that various mixtures of loans are available in the market.

The idea of choice of loans will be illustrated using two common types of student loans. The first, credit market loans (CML), impose fixed loan repayments on all students. The fixed repayments may discourage capable but disadvantaged students from acquiring higher education. Bearing fixed repayments may be too risky for disadvantaged students, considering that future incomes are uncertain at the point of taking the loan (see Del Rey and Racionero, 2010).

To overcome the under-investment in higher education, several countries, including Australia, New Zealand, Chile, Sweden and the United Kingdom, have guaranteed access to a second type of student loans: income-contingent repayment loans (ICL).

The basic feature of ICL is that repayments are based on a percentage of actual future incomes. Accordingly, students with large income realizations (discovered after they complete higher education) incur larger repayments than those with low income levels. In contrast to credit market loans (CML), ICL imply cross-subsidization among students with different future incomes: individuals with high incomes 'subsidize' those with low incomes. As a result, ICL provide 'insurance' (through risk-sharing) against the uncertainty of future incomes[1].

While other articles analyze these funding channels in isolation, Eckwert and Zilcha (2015) model the concept of choice of student loans (see also Del Rey, 2012). Their 'equilibrium with funding diversity' (FDE) allows students to choose one of two funding channels, CML or ICL. They prove that FDE reduces the risks on investments in higher education, and thereby enhances the opportunity of pursuing higher education compared to pure CML. Taking the concept of choice one step further, while they *forbid* a mixture of loans, the simple framework of this study indicates that allowing students to choose their own portfolio of both funding channels is beneficial for the economy.

---

[1] Milton Friedman (1962) was the first to mention the potential advantages of ICL for students and investors. Eckwert and Zilcha (2012) analyze alternative ICL programs that differ in the degree of risk-pooling. Chapman (2006) describes the experience in Australia, the first country to implement ICL (see also Barr and Crawford, 1998; Lleras, 2004; Nerlove, 1975; Woodhall, 1988).

The model (to be presented later in this paper) assumes that when risk-averse young individuals invest in higher education, there is an attendant amplifying of their earning potential as adults. At the point of entering the higher education system, the knowledge about their earning potential is limited to a public signal, a more or less accurate indication about their actual ability. An example to such a signal is high school achievements, which may predict future incomes to some extent. Given this uncertainty about their earning potential, individuals choose how to finance their higher education.

Introducing a basic tool of financial markets into student loans, a 'portfolio regime' (PR), grants individuals 'more choice' by allowing them to choose their own mixture of two funding channels: CML and ICL. The model is then used to examine how individuals react to being granted 'more choice'; and, is 'more choice' beneficial for the economy?

The value of choice is assessed by comparing the 'equilibrium with funding diversity' (FDE), which forbids a mixture of loans, and the PR that allows it. Eckwert and Zilcha, 2015 (proposition 2) show that the FDE is inefficient. The reason is that favorable students (students who are predicted to be successful), with high income prospects, are less likely to choose ICL. This is true because the obligation to 'subsidize' individuals with low incomes (the cross-subsidization) imposes elevated financing costs on the favorable students. In other words, ICL participants are adversely-selected.

By allowing students to choose their own portfolio of loans, some favorable students decide to take advantage of this opportunity and combine ICL in their portfolio, improving the borrowing terms of all ICL participants. Other students (who may draw extreme signals: the most favorable or the least favorable ones) finance their higher education either through one loan or the other, and may enjoy the improved borrowing terms in the ICL program. Therefore, allowing students to choose 'portfolios' can be a beneficial tool in higher education funding. Besides assessing the value of choice by comparing the FDE, which forbids a mixture of loans, and the PR that allows it, the study analyzes the funding decisions of students within the PR.

The analysis shows that within the PR, when students are allowed to choose their own portfolio, their funding decisions depend on how risk-averse the students are, and on how accurate their signals are, or how well the signals predict their earning potential. For example, the most favorable students (who are predicted to be the most successful), with the highest income prospects, may prefer the loan with the worst borrowing terms, if they are highly risk-averse or if the signals are noisy. In this case, the fixed repayments of CML may be too risky for them. In order 'to be on the safe side', they may choose ICL. While they gain 'insurance' against the uncertainty of future incomes, their borrowing terms deteriorate because they subsidize the ICL program. On the other hand, if they are not so risk-averse, they may prefer larger shares of CML in their portfolio in order to improve their borrowing terms.

To further understand how this framework works, Section 2 introduces the model in more detail. Section 3 depicts the results. While section 3.1 characterizes the funding decisions, section 3.2 assesses the value of choice of student loans. Section 4 concludes. Unless otherwise mentioned, all proofs are relegated to the Appendix.

## 2.    The Model

The following model illustrates the positive and normative implications of granting students 'more choice' in their funding decisions. The first part of the model formulates how ability and education create incomes. The second part focuses on the funding of higher education, including the government's role in the higher education system. The third part of the model describes the role of individuals and firms. Then, all these ingredients are combined to define the equilibrium.

To formulate how ability and education create incomes in the model, the roles of education are described hereby. Following the literature, education plays two standard roles. 1) Education is an investment in human capital, knowledge that pays off in terms of higher productivity. In the context of education pursuit, it is assumed that young individuals acquire education, whereas adults earn a labor income based on their education. 2) Its other role is to provide a public signal, some kind of indication, more or less accurate, on the abilities of individuals, which is assumed to assist in their decision to pursue higher education.

The model assumes that at birth, individuals (a continuous sample in the interval $[0,1]$) are randomly endowed with innate abilities, $\tilde{a}^i \in [a^1, a^2] \subset R_+$. The realization of abilities, $a^i$, is yet unknown[2].

The completion of public education (K-12) is the first step that reveals some information about abilities. As the focus of this study is higher education, assume for simplicity that public education is equal, compulsory and free of cost. In addition to a basic level of human-capital, '$A$', public education provides a signal, $y \in [y^1, y^2] \subset R_+$, an indication about abilities. Larger signals represent 'good news' because they forecast higher realizations of ability[3].

Given that at this point the information on each individual is the ability signal, denote all individuals with signal $y$ as signal group $y$, and their ability as a random variable $\tilde{a}_y$, with a mean of $\bar{a}_y = E[\tilde{a}_y]$.

After the completion of public education (K-12), young individuals in each signal group $y$ choose whether to pursue higher education ($I = 1$) or not ($I = 0$). Following a standard modeling procedure in the growth literature, investing in higher education enhances the level of human capital. If the young individuals in signal group $y$ decide not to pursue higher education, the predicted level of their human capital, $\tilde{h}_y$, remains at the basic level, $A$, provided by the public education system. In case they decide to pursue higher education, their human capital is upgraded to $A + \tilde{a}_y$.

$$(1) \quad \tilde{h}_y = \begin{cases} A & , \text{if } I = 0 \\ A + \tilde{a}_y, & \text{if } I = 1 \end{cases}.$$

To transform the predicted human capital of young individuals into predicted incomes, assume that each person supplies 1 unit of labor, and that $\omega$ is the wage rate for an effective unit of human capital. Then, the random labor income, $\omega \tilde{h}_y$, is simply the random human capital in *Eq. (1)*, multiplied by $\omega$.

---

[2] Note that a tilde denotes a random variable, whereas no tilde denotes its realization.

[3] Accordingly, assume that signals and abilities satisfy the 'Monotone Likelihood Ratio Property' (MLRP) (see Milgrom, 1981). Therefore, the ability distribution in a higher signal group 'First Degree Stochastically Dominates' the ability distribution in a lower signal group, $y' > y \Rightarrow \tilde{a}_{y'} \succ_1 \tilde{a}_y$.

The second step that reveals information about abilities is the completion of higher education. The model assumes that at this point, abilities are fully revealed. Therefore, the actual level of human capital and the actual incomes are also known at this point, and are calculated using the realization of abilities, $a^i$. The income is used for personal consumption and repaying the student loans. After formulating how ability and education create incomes, the second part of the model focuses on the funding of higher education, including the government's role in the higher education system.

To finance the cost of higher education, which is normalized to '1', two traditional types of loans are available for young individuals. 1) Credit market loans (CML), impose fixed repayment on all students, the interest rate. The interest rate is exogenously given by the international gross interest rate '$R$', assuming that the country in the model is small and open (physical capital is internationally mobile). 2) In contrast, the repayment of income-contingent repayment loans (ICL), $R\frac{\tilde{a}_y}{\bar{a}}$, differs among students.

The plug-number, $\bar{a}$, to be defined shortly, generates cross-subsidization among students with different future incomes: students with large income realizations incur larger repayments than those with low income levels. Specifically, favorable students (students who are predicted to be successful), with large predicted abilities, $\bar{a}_y > \bar{a}$, and thus high income prospects, are expected to cross-subsidize others. It is easy to notice the cross-subsidization through the expected (mean) repayments. The expected repayment of favorable students as ICL participants, $R\frac{\bar{a}_y}{\bar{a}}$, is larger than their repayment as CML participants, the interest rate, whereas the repayment of other students as ICL participants is lower than the interest rate[4].

---

[4] Recall that individuals repay their loans in adulthood after their actual abilities are revealed. Therefore, ICL participants actually cross-subsidize others if the realization of their repayment, $R\frac{a^i}{\bar{a}}$, is larger than the interest rate (the realization of their ability, $a^i$, is larger than $\bar{a}$).

To introduce numerous choices of loans to the model, a portfolio regime (PR) allows individuals to choose portfolios, or mixtures of CML and ICL, where $\theta_y \in [0,1]$ denotes the share of CML, and $1-\theta_y$ denotes the share of ICL in their portfolio. Then, the expected repayment of signal group $y$ is

$$(2) \quad \theta_y R + (1-\theta_y) R \frac{\tilde{a}_y}{\bar{a}}$$

To complete the discussion on the funding of higher education, it is left to explain how the plug-number, $\bar{a}$, is determined. It is assumed that the government designs the loan program to break even, or to balance itself, through the plug-number, $\bar{a}$.

In order to balance the loan program without being subsidized by the government, private investors, who provide the student loans, must gain the interest rate on their investment (their outside option). Note that while providing a loan to each individual separately may be risky for them, because abilities are yet unknown at the point of taking the loan, uncertainty does not exist in the economy level. Students, as a group, are non-risky, because the distribution of their abilities is assumed to be known. As a result, the investors, who buy a non-risky asset, are willing to gain the interest rate.

Therefore, in order to balance the loan program the government equates the expected repayment of student loans *across all signal groups* to the interest rate,

$$(3) \quad E\left[\theta_y R + (1-\theta_y) R \frac{\bar{a}_y}{\bar{a}}\right] = R.$$

Rearranging Eq. (3), $\bar{a}$ is a plug-number calculated by the government as a weighted mean ability of ICL participants (with $\theta_y < 1$),

$$(4) \quad \bar{a} = \frac{E\left[(1-\theta_y)\bar{a}_y\right]}{E\left[1-\theta_y\right]}.$$

Eq. (4) illustrates how the flow of students to and from the ICL program affects the borrowing terms of other students. For example, if high-signal groups, with $\bar{a}_y > \bar{a}$, join the ICL program, then the borrowing terms improve ($\bar{a}$ rises, which means that the expected repayment of *all* ICL participants, $R\frac{\bar{a}_y}{\bar{a}}$, declines). Therefore, the choice of loans by each signal group may affect other students. After describing all the various choices of higher education funding, the third part of the model presents the role of individuals and firms. Allowing the use of portfolios as an addition to traditional loans, how do individuals actually choose their preferred mix of loans?

The model assumes that individuals choose their mix of loans so as to gain utility from personal consumption of their labor income after repaying the loans, $c_y$. In other words, students choose the share of CMLs, $\theta_y$, in their portfolio that maximizes their expected utility from consumption $u(\tilde{c}_y): R_+ \to R$ (see definitions 1-2 in the Appendix for formulation and standard assumptions).

Note that to avoid a trivial solution, it is necessary to ensure that investment in ICL is beneficial for some signal group $y$. In other words, it is ensured that the expected income of some signal group $y$ is larger than its expected ICL repayment, $\tilde{a}_y\left(\omega - \dfrac{R}{\bar{a}}\right) > 0$ by assuming that $\bar{a}\omega > R$. It is easy to notice, though, that if investment in ICL is beneficial for *some* signal group, then it is beneficial for *all* signal groups.

This feature of ICL that all individuals invest in higher education does not exist in reality, because higher education institutions set admission standards; individuals with low signals are not accepted to higher education programs. Therefore, incorporating admission standards would achieve a more realistic model.

However, admission standards are excluded from the model to keep it simple, and because admission standards have no bearing on the qualitative results. Following the above discussion concerning individuals, the last part of the model introduces firms.

Upon reaching adulthood consumers purchase a consumption product from competitive identical firms. The firms produce the product using the stock of human capital in the economy, $H = A + E[a^i]$, which is calculated using the distribution of abilities. The other input of production is physical capital, *K*, that fully depreciates in the production process.

The production function, $F(K,H)$, exhibits standard assumptions: returns to scale is constant; the marginal products are positive and decreasing; firms hire physical capital and human capital until their marginal products equal the interest rate and the wage rate, respectively,

$$(5) \quad \begin{aligned} F_K\left(\frac{K}{H},1\right) &= R \\ F_L\left(\frac{K}{H},1\right) &= \omega \end{aligned}$$

Combining the ingredients of the model, the firms, the individuals, and the government, the equilibrium can be solved.

To define the equilibrium, note that *Eq. (5)* solves the physical capital and the wage rate, given the stock of human capital and the (international gross) interest rate[5]; The government chooses the weighted mean ability of ICL participants, $\bar{a}$, so that the loan program breaks even according to *Eq. (4)*; Each individual maximizes the expected utility in order to choose the share of CML, $\theta_y$, as a function of $\bar{a}$. Substituting $\theta_y(\bar{a})$ in *Eq. (4)* solves the equilibrium.

**Definition of Equilibrium:**
*Given the international gross interest rate, the distribution of abilities, the distribution of signals, and the stock of human capital, $H$, the equilibrium consists of a vector $(\omega, K, H) \in R_+^3$, and a share of CML, $\theta_y$, for each signal group, such that*

i.   Firms: The wage rate and the physical capital satisfy Eq. (5);
ii.  Government: The loan program breaks even by $\bar{a}$ according to Eq. (4); and
iii. Individuals: The share of CML, $\theta_y$, maximizes their expected utility, $u(\tilde{c}_y)$. 。

---

[5] The first equality implies that the interest rate, $R$, uniquely determines the ratio $\frac{K}{H}$ and, thus, $K$. Then, the wage rate, $\omega$, is uniquely determined by substituting $\frac{K}{H}$ in the second equality.

After the introduction of the model, the following section reports the results.

## 3. Results

The remainder of the article explores how the various choices of funding available in the market affect students. It is, then, established that granting students 'more choice' by allowing them to choose portfolios of loans is beneficial for the economy.

Considering the various choices of loans, there are attendant questions. When offered the choice to mix loans, do students practically use this opportunity or prefer to finance their higher education either through one type of loan or the other? And if some students indeed mix their loans, what are the motivations and features of their decisions? The answer is not obvious, because it is a choice, not an obligation.

Some students (also referred to as the portfolio set) indeed use this opportunity and choose a portfolio of loans. To confirm that, let signal group $\hat{y}$ be indifferent between one loan (ICL-only, $\theta_y = 0$) or the other (CML-only, $\theta_y = 1$) (see definition 3 on the Appendix), and assume that $\hat{y}$ lies in the signal distribution,

**Assumption** 1: *Signal group $\hat{y}$ lies in the signal distribution, i.e., $\hat{y} \in (y^1, y^2)$.* 。

Then, it is easy to verify that the individuals from signal group $\hat{y}$ who are indifferent between the two loans, strictly prefer to mix the two loans, because of the risk-aversion. By diversifying the loans, they gain some insurance against the uncertainty of future incomes. But not only signal group $\hat{y}$ prefers portfolios. Continuity implies that signal groups in a sufficiently small neighborhood around $\hat{y}$ also prefer to combine the two loans. Therefore,

**Proposition 1** *(existence of the portfolio set)*
*Under assumption 1, the portfolio set is not empty.* 。

In the sequel, let $y'$ and $y''$ be the cutoff signals that separate the portfolio set from individuals who prefer ICL-only or CML-only, respectively (see definition 4 on the Appendix).

To further understand the choice of student loans, after realizing that some students indeed mix their loans, the second attendant question is what are the motivations and features of their decision to have a portfolio of loans? To answer this question, recall that students choose the share of CMLs, $\theta_y$, in their portfolio that maximizes their expected utility from consumption, $u(\tilde{c}_y)$. Deriving the expected utility by $\theta_y$ yields[6]

$$(6) \quad \frac{\partial E\left[u(\tilde{c}_y)\right]}{\partial \theta_y} = \frac{R}{\bar{a}}\left((\bar{a}_y - \bar{a})E\left[u'(\tilde{c}_y)\right] + cov(\bar{a}_y, u'(\tilde{c}_y))\right).$$

The first-order condition recognizes two motives that are common in the analysis of financial markets. The first component in *Eq. (6)* reflects the desire to reduce expected repayments. Its sign depends on the signal's magnitude. Recall that in the ICL program, high-signal groups, with $\bar{a}_y > \bar{a}$, cross-subsidize low-signal groups, with $\bar{a}_y \leq \bar{a}$, because their expected repayment are larger than the interest rate. Thus, if students' only concern was their expected repayment, then high-signal students would prefer CML-only, $\theta_y = 1$, with fixed repayments given by the interest rate, while low-signal students would prefer the chance to be subsidized by ICL-only, $\theta_y = 0$.

The second component in *Eq. (6)*, $cov(\bar{a}_y, u'(\tilde{c}_y))$, has a negative sign, which guides students to choose ICL as an insurance tool[7]. Based on these motives, it is easy to establish that low-signal groups, with $\bar{a}_y \leq \bar{a}$, behave differently from the rest of the students.

Low-signal students (with low income prospects) finance their higher education entirely through the ICL program for two reasons. The cross-subsidization provides them relatively improved borrowing terms and, at the same time, risk-sharing against the uncertainty of future incomes[8].

---

[6] $\theta_y$ is unique because the utility function is concave and the consumption is linear w.r.t $\theta_y$.

[7] The sign is negative, because the utility function is concave, and $\partial \bar{c}_y / \partial \bar{a}_y = \omega - \frac{(1-\theta_y)R}{\bar{a}} > 0$. Note that in case students were risk-neutral, the equilibrium would collapse to CML-only for all $y$.

**Proposition 2** *(existence of the ICL-only set):*
*Assume that $y$ satisfies $\bar{a}_y \leq \bar{a}$. Then, $\theta_y = 0$.* ∘

While low-signal groups prefer ICL-only, and thus the ICL-only set always exists, the funding decisions differ among high-signal groups. Some high-signal groups finance their higher education through one loan only, whereas the rest diversify their loans[9],

**Corollary 1** *(the portfolio set consists of high-signal groups):*
*The portfolio set exhibits $\bar{a}_y > \bar{a}$.* ∘

The motivation for the heterogeneous funding decisions is that high-signal groups face a tradeoff between expected repayment and risk. ICL provide them 'insurance' against the uncertainty of future incomes and, at the same time, less favorable borrowing terms than CML (being obliged to cross-subsidize others). These contradicting incentives produce a variety of choices of high-signal groups, which hinge on their specific signal and utility function.

Assume, for example, that the utility function exhibits 'constant absolute risk-aversion' (CARA), the absolute risk-aversion is constant throughout all signal values (see definition 2 on the Appendix). In this case, the choice of loans in high-signal groups depends on the specific signal, because the signal determines the ICL borrowing terms. In the ICL program, students with larger signals, and thereby higher income prospects, are 'penalized' with larger expected repayments, $R\frac{\bar{a}_y}{\bar{a}}$, than students with lower signals. As a result, they are encouraged to shift to the CML program in order to improve their borrowing terms[10].

---

[8] They actually prefer a negative share of CML, but are restricted to choose $\theta_y \in [0,1]$. Note that even signal groups $y = \bar{a}$, who obtain identical expected repayments under CML or ICL, strictly prefer ICL-only because of the risk-aversion.

[9] It is easily verified that $\bar{a}_y > \bar{a}$ is a necessary condition for an interior solution in *Eq. (6)*. As low-signal groups prefer ICL-only, it is straightforward that the portfolio set consists of high-signal ones.

This effect is reinforced under 'constant relative risk-aversion' (CRRA) utility function, given its well-known feature that individuals with larger signals, and thereby higher income prospects, are *less* risk-averse (see definition 2 on the Appendix). Less risk-averse individuals prefer a larger share of CML in their portfolio, or improved borrowing terms at the expense of the risk-sharing advantages of the ICL program. Therefore, under CARA or CRRA utility functions, high-signal groups use larger shares of CML as their signal rises.

**Proposition 3** *(high-signal groups: heterogeneity):*
*Under CRRA or CARA utility functions, $\theta_y$ is weakly increasing in $y$.* ∘

Fig. 1 illustrates the choice of loans with CRRA or CARA utility functions (propositions 1-3). The solid line denotes a PR. For comparison, the dashed line denotes Eckwert and Zilcha (2015)'s FDE that forbids portfolios. Fig. 1 depicts that students use larger shares of CML as the signal rises. Low-signal groups finance their higher education entirely through the ICL program. In contrast, for high-signal groups the ICL program has a cost in terms of expected consumption. As a result, a set of them prefers to diversify their loans (including signal group $\hat{y}$). Figure 1 illuminates that students with signals at either extreme (the most favorable or the least favorable ones) select one loan only; that is, their decisions are identical in the PR and the FDE. In contrast, students with intermediate signals, not too high and not too low, choose a mix of the two loans.

---

[10] The price effect consists of a substitution effect (ICL become more expensive, because the expected repayments rise) and a negative income effect. According to Landsberger and Meilijson (1989, proposition 2), the substitution effect offsets the income effect.

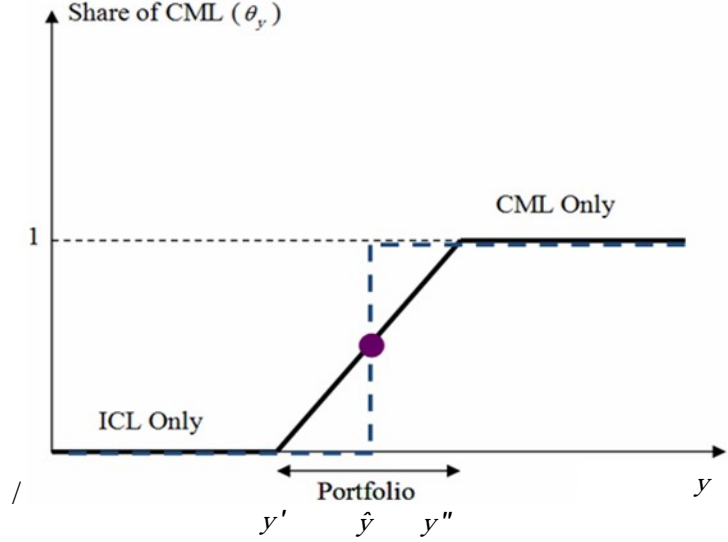

**Fig. 1.** Illustration of funding decisions under CRRA or CARA utility functions. The x-axis denotes the signal, $y$, and the y-axis denotes the CML share in student loans $\theta_y \in [0,1]$ ($\theta_y = 0$ for ICL-only; $\theta_y = 1$ for CML-only). The solid line (linear for simplicity of illustration) represents the PR. For comparison, the dashed line denotes Eckwert and Zilcha (2015)'s FDE that forbids portfolios. Fig. 1 illustrates that the CML share is increasing under CRRA or CARA utility functions. Moreover, intermediate-signal students strictly prefer portfolios (see propositions 1-3).

To further study the motivations of high-signal groups, and specifically how their choice of loans hinges on their specific signal and utility function, take the commonly used quadratic utility function, $u(\tilde{c}) = \alpha\tilde{c} - \frac{1}{2}\beta\tilde{c}^2$, with standard parameters $\alpha > 0$, $\beta > 0$, and $\frac{\alpha}{\beta} > \tilde{c}_y$, $\forall \tilde{c}_y$ (see definition 2 on the Appendix). The quadratic utility function provides a tractable closed-form solution for the choice of loans.

The following discussion establishes that in the case of quadratic utility functions the choice of loans by high-signal groups defers from the previously studied (CRRA and CARA) utility functions. Assumption 2 simplifies the analysis,

**Assumption 2**: $\tilde{a}_y = y + \tilde{\varepsilon}$, and $\tilde{\varepsilon} \sim (0, \sigma^2)$.

According to assumption 2, the signal reflects the expected ability in signal group $y$, $\bar{a}_y = E[\tilde{a}_y] = y$. The variance, $\sigma^2 > 0$, measures the signal noise (or the signal quality). In other words, as the variance increases, the prediction of ability is less accurate.

Using assumption 2, the first-order condition (*Eq. (6)*), and the consumption, $\tilde{c}_y$ (*Eq. (8)* on the Appendix), the quadratic utility function provides a tractable closed-form solution for the choice of loans (see definition 2 and lemma 1 on the Appendix).

$$(7) \quad \theta_y = \frac{k\bar{a}(y-\bar{a})}{R\left((y-\bar{a})^2 + \sigma^2\right)} - \left(\frac{\bar{a}\omega - R}{R}\right).$$

The closed-form solution reveals surprising choices of loans by high-signal groups. In contrast to the CRRA and CARA utility functions, a larger signal does not necessarily encourage high-signal groups to use larger shares of CML.

On the one hand, the larger the signal is, the less favorable are the ICL borrowing terms, which pushes high-signal groups to the CML. On the other hand, a well-known feature of quadratic utility functions is that individuals with larger signals, who are thereby expected to earn a larger labor income, are *more* risk-averse (see definition 2 on the Appendix). The increasing risk aversion generates a contradicting effect on the choice of loans. More risk-averse individuals are willing to forgo consumption (or suffer deteriorated borrowing terms) in order to acquire ICL as a risk-sharing tool.

These two contradicting effects derive the funding decisions summarized in Fig. 2.

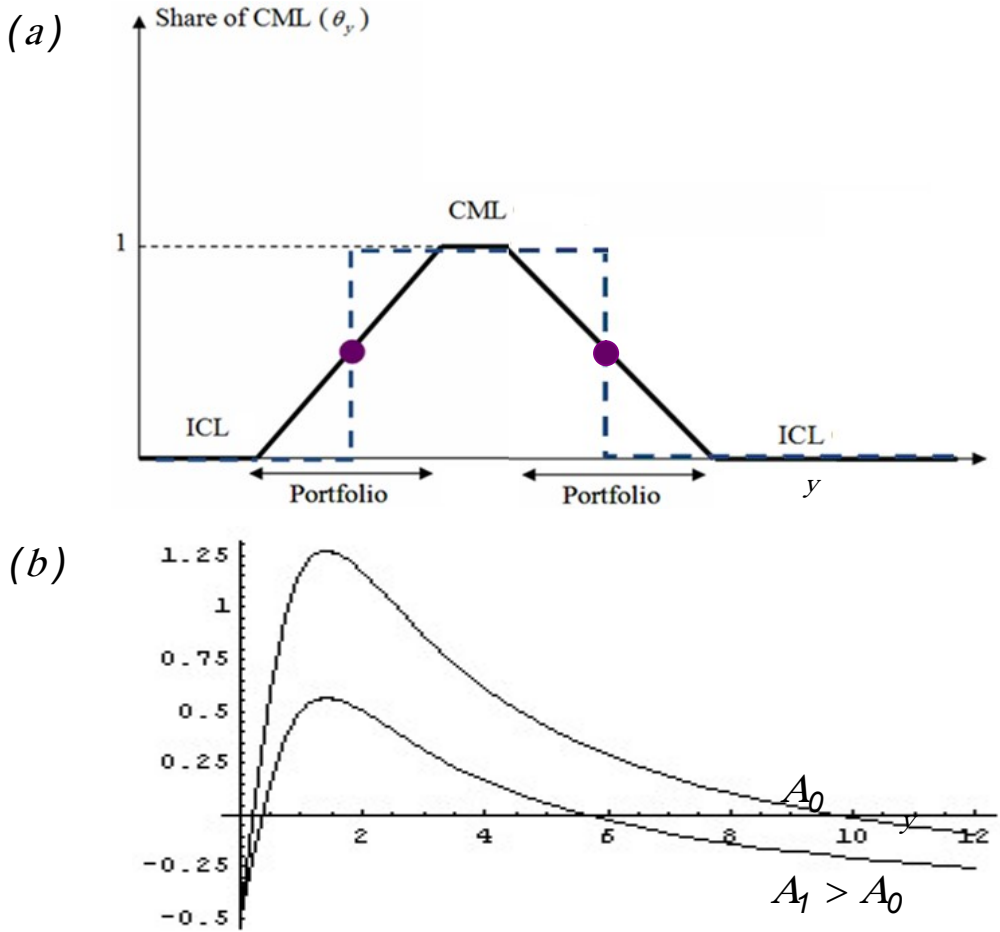

**Fig. 2.** Funding decisions under quadratic utility functions. Notations are similar to Fig. 1. Fig. 2(a) illustrates that students with extreme signals choose ICL-only (including the most favorable students), whereas the rest of the students choose CML-only or portfolios. For simplicity of illustration, lines are linear. Fig. 2(b) depicts two numerical examples of *Eq. (7)* to illustrate the comparative static. The specific parameters of Fig. 2(b) are available on request (see further details in the sequel). The lower Fig. takes a higher basic level of human capital, $A_1 > A_0$, which makes students more affluent, and thereby more risk-averse, and, as a result, they lower their CML share. The numerical examples also demonstrate that several students actually prefer $\theta_y \notin [0,1]$, but are restricted to choose $\theta_y \in [0,1]$.

Fig. 2(a) depicts that students with low signals increase their CML share as the signal rises in order to reduce their expected repayment. This behavior is qualitatively similar to the cases of CRRA and CARA utility functions.

However, for sufficiently large signals, the opposite case occurs. The increasing absolute risk-aversion (ARA, in short) becomes the dominant factor. As the signal rises, students increase the ICL share because they become more risk-averse. As the most favorable students, with the highest income prospects, are also the most risk-averse, they prefer ICL-only. They choose ICL-only in order to gain risk-sharing, but at the same time they pay a high cost in their borrowing terms (because they are expected to subsidize the ICL program). While students with extreme signals

(including the most favorable students) participate solely in the ICL program, the rest of the students choose portfolios or CML-only. This result, which is specific to quadratic utility functions, reflects the concavity of the CML share.

**Proposition 4** *(High-signal groups: heterogeneity):*
*Under quadratic utility functions, $\theta_y$ is a concave function of $y$.* ∘

To further explore the motivations and features of the funding decisions aside from the signal itself, the following discussion focuses on two other circumstances that determine the choice of loans. 1) How risk-averse the students are. Under quadratic utility functions, the ARA can rise directly (if $\beta/\alpha$ rises) or indirectly (if the basic level of human-capital, $A$, grows, which makes students more affluent, and thereby more risk-averse). 2) Funding decisions also depend on how accurate the signals are, or how well the signals predict the earning potential. Recall that the signal accuracy is measured by the variance of the random ability, $\sigma^2$, see assumption 2. As the variance increases, the prediction of ability is less accurate. Corollary 2 summarizes the effect of the ARA and the signal quality on the choice of loans.

**Corollary 2** *(comparative static: ARA, and signal quality):*
*Under quadratic utility functions, $\theta_y$ is decreasing in $\beta/\alpha$, $A$, and $\sigma^2$.* ∘

According to corollary 2, when students are more risk-averse or when the signals are noisier (less accurate), students augment their ICL share as a risk-sharing tool[11]. To demonstrate this argument, fig. 2(b) uses two numerical examples of *Eq. (11)*[12]. In the upper Fig., the baseline, some students choose portfolios, while other students choose CML-only or ICL-only. The lower Fig. takes a higher ARA (particularly, a higher basic level of human capital, $A_1 > A_0$). As a result of the higher ARA, the risk-sharing considerations lead students to lower their CML share to the extent that, in this case, the set of students who choose CML-only is eliminated completely.

---

[11] In a similar model, Hatsor (2015) examines the role of information in students' funding decisions, and finds that more information (more accurate signals) can be harmful to the economy.
[12] The specific parameters of Fig. 2(b) are available on request (see further details in the sequel).

Note that sufficiently large ARA or sufficiently noisy signals may lead to a trivial choice of loans, where all students choose ICL-only. To avoid this trivial solution, it is assumed in the case of quadratic utility functions that the parameters of ARA and the signal noise are sufficiently small.

**Assumption 3**: Under quadratic utility functions, assume that $\sigma^2 < \left(\frac{k\bar{a}}{2(\bar{a}\omega - R)}\right)^2$. 。

A more binding assumption on the upper bound of the signal distribution, $y^2$, in addition to small ARA and small signal noise, is necessary to ensure that some students choose CML-only[13].

**Assumption 4**: Under quadratic utility functions, assume that $\sigma^2 \leq \left(\frac{k}{2\omega}\right)^2$ and $y^2 \geq y''_1$. 。

Proposition 5 uses assumption 3-4 to assure that funding decisions are heterogeneous. That is, students choose various types of loans.

**Proposition 5:** *(existence of various types of loans):*
*Under quadratic utility functions,*
  *(a) If assumption 3 holds, then the ICL-only set and the portfolio set exist.*
  *(b) If assumption 4 holds, then all sets exist: ICL-only, portfolios, and CML-only*。

Fig. 2(b) illustrates the results of Proposition 5 (the specific parameters are available on request) using two numerical examples. In the upper Fig., the baseline, assumption 4 holds, because the ARA and the signal noise are sufficiently small, and thus students choose various types of loans. The lower Fig. differs from the baseline in one aspect that breaks assumption 4. It assumes a larger level of basic human capital. As a result, students are more risk-averse, and thus their loans always contain a certain share of ICL. In other words, the CML-only set, where all students choose CML-only, is eliminated completely in this case. After characterizing the choice of loans, the article examines whether 'more choice' by allowing the use of portfolios as an addition to traditional loans is beneficial for the economy.

---

[13] Recall that $y''_1$ denotes the lower cutoff signal between the portfolio set and the CML-only set (see definition 4 on the Appendix). Note that assumption 1 is more general than assumptions 3-4 and avoids trivial solutions for all utility functions.

*3.2. The value of choice of student loans*

Given that some students actually prefer to mix their loans, the remainder of the article examines how their decisions affect the economy. This is an importantly practical question for governments that consider whether to grant students 'more choice' in their funding decisions.

To assess the value of choice, a 'portfolio regime' (PR) that permits students to choose their own portfolio of loans is compared to 'equilibrium with funding diversity' (FDE), which forbids a mixture of loans. The following discussion establishes that granting students 'more choice' by allowing the use of portfolios is beneficial for the economy.

In a PR, some high-signal groups decide to use the opportunity and mix the two loans. They clearly benefit from the PR because they *voluntarily* choose a portfolio. However, 'more choice' may be valuable not only for these students, who prefer to use a portfolio of loans, but also for other students.

The rest of the students do not change their funding decisions and still finance their higher education either through one loan or the other. With the same fixed repayment, students who remain entirely in the CML market are indifferent between the two regimes. In contrast, students who remain entirely in the ICL program are affected by the decision of other students to mix loans. Specifically, participation of high-signal groups in the ICL program improves the borrowing terms (because they subsidize the ICL program, see *Eq.* (4) [14]).

---

[14] Formally, the expected repayment, $R\frac{\bar{a}_y}{\bar{a}}$, declines because high-signal groups increase the weighted mean ability of ICL participants, $\bar{a}$.

By allowing students to choose their own portfolio of loans, some high-signal groups, who use CML-only in the FDE, indeed enter the ICL program, or reduce their share of CML, $\theta_y$ (see Fig. 1, to the right of $\hat{y}$). When they decide to mix loans, they improve the borrowing terms of all ICL participants.

Note that other high-signal groups, with lower signals, who use ICL-only in the FDE, opt out of the ICL program (see Fig. 1, to the left of $\hat{y}$), slightly worsening the borrowing terms. Their effect on the borrowing terms is minor, though, because of their lower signals.

Therefore, considering the overall effect of a PR, ICL participants enjoy the improved borrowing terms in the ICL program. Given that all students are weakly better off in the PR, 'more choice' by allowing the use of portfolios is beneficial for the economy[15].

**Proposition 6** *(more choice in student loans is desirable):*

*Assume that the utility function is CRRA or CARA. Or assume that the utility function is quadratic and $y^2 \leq \hat{y}_2$. Then, a PR Pareto dominates a FDE.* ∘

Proposition 6 illuminates another advantage of the PR. Eckwert and Zilcha, 2015, (see Proposition 2 there) point out that the FDE suffers from adverse selection, high-signal groups under-invest in ICL. This is true because the obligation to 'subsidize' individuals with low incomes (the cross-subsidization) imposes elevated financing costs on students with high incomes. However, high-signal groups do not consider their (expected) positive effect on *all* ICL participants, and thereby depart from the ICL program, worsening the borrowing terms for all students[16]. In the PR, by allowing students to choose their own portfolio of loans, high-signal groups are encouraged to combine ICL in their choice of loans, alleviating the adverse selection, or their under-investment in ICL. Therefore, 'more choice' by allowing the use of portfolios as an addition to traditional loans can be a beneficial tool in higher education funding.

---

[15] If the utility function is quadratic and the upper bound of the signal distribution is sufficiently small, the proof strengthens. The borrowing terms of ICL participants further improve because another set of high-signal students enters the ICL program (see Fig. 2(a) for illustration). Note that if assumption 1 does not hold, and specifically $y^2 \in (y', \hat{y})$, then by definition, under a FDE all students choose ICL-only. Shifting to a PR, high-signal groups leave the ICL program, which deteriorates the ICL borrowing terms. A formal proof is available on request.

## 4. Conclusion and policy implications

Alternative solutions to traditional policies of funding may offer viable gains to governments and individuals. Diversity in governmental funding channels is absent in most areas of the economic policy such as R&D funding or pensions. This article establishes that students benefit from a government policy that grants them 'more choice' in their funding decisions, because some individuals, then, prefer mixtures of loans, improving the borrowing terms of other students. Policies that allow the freedom to choose a diversified portfolio, a well-known tool in financial markets, may improve risk-sharing not only in the case of higher education funding studied here but also in other areas of the economic policy.

## 5. Appendix

**Definition 1** *(consumption):*
Consumption in adulthood conditional on the signal $c_y$, given the human capital *(Eq. (1))*, equals

$$(8) \quad \tilde{c}_y = \begin{cases} A\omega & , \text{ if } I = 0 \\ \underbrace{(A + \tilde{a}_y)\omega}_{\text{income}} - \underbrace{\left[ \overbrace{\theta_y R}^{\text{CML}} + (1 - \theta_y) R \overbrace{\frac{\tilde{a}_y}{\bar{a}}}^{\text{ICL}} \right]}_{\text{repayment obligation}} & , \text{ if } I = 1 \end{cases}.$$

**Definition 2** *(utility):*
It is assumed that utility is von Neumann–Morgenstern, twice differentiable, strictly increasing and concave. Several utility functions are used: constant relative risk-aversion (CRRA), constant absolute risk-aversion (CARA) and quadratic:

---

[16] Students take $\bar{a}$, the weighted mean ability of all ICL participants, as given. However, high-signal ICL participants increase $\bar{a}$, which reduces the expected ICL repayments, $R\frac{\bar{a}_y}{\bar{a}}$.

a. *Utility functions* $u(\tilde{c}) = \dfrac{\tilde{c}^{1-\gamma}}{1-\gamma}$ exhibit CRRA. The ARA, $-\dfrac{u''}{u'} = \dfrac{\gamma}{\tilde{c}}$, is decreasing as income rises.

b. *Utility functions* $u(\tilde{c}) = 1 - e^{-\lambda \tilde{c}}$ exhibit CARA. The ARA equals $\lambda$.

c. *Utility functions* $u(\tilde{c}) = \alpha \tilde{c} - \dfrac{1}{2}\beta \tilde{c}^2$ are quadratic. The ARA, $\dfrac{\beta}{\alpha - \beta \tilde{c}}$, is increasing as income rises.

Because the utility function is strictly increasing and concave, $\alpha > 0$, $\beta > 0$, and $\dfrac{\alpha}{\beta} > \tilde{c}_y$, $\forall \tilde{c}_y$.

Deriving the expected utility by $\theta_y$ obtains the interior solution:

$$(9) \qquad \frac{\partial E\left[u(\tilde{c}_y)\right]}{\partial \theta_y} = kE\left[\tilde{a}_y - \bar{a}\right] - \left(\omega - \frac{R}{\bar{a}}(1-\theta_y)\right)E\left[(\tilde{a}_y - \bar{a})^2\right] = 0$$

where $k = \dfrac{\alpha}{\beta} - \left((A+\bar{a})\omega - R\right) > 0.$

Rearranging *Eq. (9)*, the CML share is solved explicitly. ∘

**Lemma 1**: If $y > \bar{a}$, then $k > 0$.

Proof: The expected consumption (see *Eq.(8)*) is given by

$$(10) \quad \begin{aligned} E(\tilde{c}_y) &= A\omega + E(\tilde{a}_y)\left(\omega - (1-\theta_y)\frac{R}{\bar{a}}\right) - \theta_y R \\ &> A\omega + \bar{a}\left(\omega - (1-\theta_y)\frac{R}{\bar{a}}\right) - \theta_y R \\ &= (A+\bar{a})\omega - R \end{aligned}$$

The inequality derives from $y > \bar{a}$. Note that $\dfrac{\alpha}{\beta} > E(\tilde{c}_y)$ (recall definition 2). Then, $\dfrac{\alpha}{\beta} > (A+\bar{a})\omega - R$. Therefore, $k > 0$. ∘

**Definition 3**: Denote by $\tilde{c}_{y,0}$ and $\tilde{c}_{y,1}$ the random consumption with ICL-only ($\theta_y = 0$) or CML-only ($\theta_y = 1$). *The cutoff signal group $\hat{y}$ is indifferent between ICL-only and CML-only*

$$(11) \qquad E\left[u(\tilde{c}_{\hat{y},0})\right] = E\left[u(\tilde{c}_{\hat{y},1})\right] > u(A\omega).$$

*If two cutoff signal groups $\hat{y}$ exist, I index them as $\hat{y}_1 < \hat{y}_2$*. ∘

The equality implies that the CML program co-exists alongside the ICL program. The assumption that investment in higher education is profitable, $\bar{a}\omega > R$, ensures that the ICL program does not break down. Thus, the inequality holds.

**Definition 4**: *The cutoff signal groups $y'$ and $y''$ are indifferent between a portfolio and ICL (CML)-only, respectively*:

$$(12) \qquad E\left[u(\tilde{c}_{y',\theta})\right] = E\left[u(\tilde{c}_{y',0})\right] \text{ and}$$

$$E\left[u(\tilde{c}_{y'',\theta})\right] = E\left[u(\tilde{c}_{y'',1})\right].$$

*If two cutoff signal groups $y''$ exist, I index them as $y''_1 < y''_2$.* °

**Proof of proposition 1**:

It is easy to verify that the consumption of students who diversify their loans is a linear combination of $\tilde{c}_{y,0}$ and $\tilde{c}_{y,1}$. That is, $\tilde{c}_{y,\theta} = \theta \tilde{c}_{y,1} + (1-\theta)\tilde{c}_{y,0}$, for all $\theta \in (0,1)$. The concave utility function satisfies the following:

(13) $\quad E[u(\tilde{c}_{y,\theta})] > \theta E[u(\tilde{c}_{y,0})] + (1-\theta) E[u(\tilde{c}_{y,1})]$.

Inequality (13) can be rewritten as

(14) $\quad E[u(\tilde{c}_{y,\theta})] > \theta \left( E[u(\tilde{c}_{y,0})] - E[u(\tilde{c}_{y,1})] \right) + E[u(\tilde{c}_{y,1})]$;

or

$E[u(\tilde{c}_{y,\theta})] > (1-\theta) \left( E[u(\tilde{c}_{y,1})] - E[u(\tilde{c}_{y,0})] \right) + E[u(\tilde{c}_{y,0})]$.

Now, substituting *Eq. (11)*, the *cutoff signals* $\hat{y}$ satisfy

(15) $\quad E[u(\tilde{c}_{\hat{y},\theta})] > E[u(\tilde{c}_{\hat{y},1})]$

and

$E[u(\tilde{c}_{\hat{y},\theta})] > E[u(\tilde{c}_{\hat{y},0})]$ for all $\theta \in (0,1)$. °

**Proof of proposition 2**:

The two terms of *Eq. (6)* have a negative sign, because $\bar{a}_y \leq \bar{a}$. °

**Proof of proposition 4**:

To simplify the presentation, I define $A' = \dfrac{k\bar{a}}{R}$ and $C = \dfrac{\bar{a}\omega - R}{R}$. Substituting $A'$ and $C$ in $\theta_y$, given by *Eq. (7)*, and rearranging, the four cutoff signals between a portfolio and ICL (CML)-only are

$y'_{1,2} - \bar{a} = \dfrac{A'}{2C} \pm \sqrt{\left(\dfrac{A'}{2C}\right)^2 - \sigma^2}$ and $y''_{1,2} - \bar{a} = \dfrac{A'}{2(C+1)} \pm \sqrt{\left(\dfrac{A'}{2(C+1)}\right)^2 - \sigma^2}$, respectively. $\theta_y$ is positive if $-C(y-\bar{a})^2 + A'(y-\bar{a}) - C\sigma^2 > 0$. Concavity follows from $C > 0$ (recall the assumption that investment in higher education is profitable, $\bar{a}\omega > R$). The maximum level of $\theta_y$ derives from $\dfrac{\partial \theta_y}{\partial (y-\bar{a})} = \dfrac{-A\left((y-\bar{a})^2 - \sigma^2\right)}{\left((y-\bar{a})^2 + \sigma^2\right)^2} = 0$. That is, $y - \bar{a} = \pm \sigma$. °

**Proof of proposition 5**:

(a) Recall that $\theta_y$, given by *Eq. (7)*, is positive if $-C(y-\bar{a})^2 + A'(y-\bar{a}) - C\sigma^2 > 0$. Assumption 3 ($A'^2 \geq 4\sigma^2 C^2$) ensures that there are two positive roots to this quadratic equation, given by

$y'_1 - \bar{a} < y'_2 - \bar{a}$. $\theta_y$ assumes positive values between the roots because it is concave. Moreover, low-signal groups always choose $\theta_y = 0$ (recall proposition 2).

(b) $\theta_y$ equals '1' (recall that it cannot exceed 1) if $-(C+1)(y-\bar{a})^2 + A'(y-\bar{a}) - \sigma^2(C+1) \geq 0$. Assumption 4 ($A'^2 \geq 4\sigma^2(C+1)^2$) ensures that there is a single root or two positive roots to this quadratic equation, given by $y''_{1,2} - \bar{a}$. $\theta_y = 1$ between the roots because it is concave.

**Proof of proposition 6**:
In section 00, I verify that the CML-only set and the portfolio set are not damaged by the PR. To complete the proof, I establish that the ICL-only set is better off in the PR. I prove that $\bar{a}$ increases, which lowers the expected repayments of ICL participants, $R\frac{\bar{a}_y}{\bar{a}}$. Shifting from a FDE to a PR, the change in $\bar{a}$ (4) (the weighted mean ability of ICL participants ($\theta_y < 1$)) is measured by

$$\begin{aligned}
\Delta \bar{a} &= \frac{E^{PR}\left[(1-\theta_y)\bar{a}_y\right]}{E^{PR}\left[1-\theta_y\right]} - E^{FDE}\left[\bar{a}_y | \theta_y = 0\right] \\
&> E^{PR}\left[(1-\theta_y)\bar{a}_y\right] - E^{FDE}\left[\bar{a}_y | \theta_y = 0\right]
\end{aligned} \quad (16)$$

The equality derives because in a FDE, $\bar{a} = E\left[\bar{a}_y | \theta_y = 0\right]$. The inequality derives from $E^{PR}\left[1-\theta_y\right] < 1$.

First, assume that $y^2 > y''_1$. If the utility function exhibits CRRA or CARA, then $\theta_y$ is an increasing function of $y$ (recall proposition 3). In this case, shifting from a FDE to a PR, the portfolio set changes $\bar{a}$. Individuals $[y', \hat{y}]$ partially exit the ICL program, thereby their contribution to $\bar{a}$ declines by $E^{\hat{y}}_{y'}\left[\theta_y \bar{a}_y\right]$. In contrast, individuals $[\hat{y}, y'']$ partially enter the ICL program, thereby their contribution to $\bar{a}$ increases by $E^{y''}_{\hat{y}}\left[(1-\theta_y)\bar{a}_y\right]$; that is,

$$\begin{aligned}
\Delta \bar{a} &> E^{y''}_{\hat{y}}\left[(1-\theta_y)\bar{a}_y\right] - E^{\hat{y}}_{y'}\left[\theta_y \bar{a}_y\right] \\
&= E^{y''}_{\hat{y}}\left[\bar{a}_y\right] - E^{y''}_{y'}\left[\theta_y \bar{a}_y\right] \\
&> 0
\end{aligned} \quad (17)$$

The last inequality derives from assumption 2 and $\theta_y < 1$. Now, if $y^2 < y''_1$ (and recall that assumption 1 holds), then the set $[\hat{y}, y'']$ becomes $\left[\hat{y}, y^2\right]$ and the proof remains.

Second, assume that the utility function is quadratic. If $y^2 \leq y''_2$, then $\theta_y$ is an increasing function of $y$ (see proposition 4). Therefore, the proof is similar, where $y'_1$, $\hat{y}_1$ and $y''_1$ correspond to $y'$, $\hat{y}$ and $y''$. If $y''_2 < y^2 \leq \hat{y}_2$, then an additional set $\left[y''_2, y^2\right]$ partially enters the ICL program and further increases $\bar{a}$. Thus, $\Delta \bar{a}$ is even larger.

# References

Barr, N., Crawford, I., 1998. Funding higher education in an age of expansion. Education Economics 6, 45-70.

Barro, R., 1998. The Determinants of Economic Growth. MIT Press, Cambridge, MA.

Bassanini, S., Scarpenta, S., 2001. Does human capital matter for growth in OECD countries? Evidence form pooled mean-group estimates. OECD Economics Department, WP # 282.

Chapman, B., 2006. Income contingent loans for higher education: International reforms. In: Hanushek, E., Welch, F. (Eds.), Handbook of the Economics of Education, Vol. 2, North Holland, Amsterdam, 1435-1500.

Checchi, D., 2006. The Economics of Education, Cambridge University Press, Cambridge.

Del Rey, E., 2012. Deferring higher education fees without relying on contributions from non-students. Education Economics 20(5), 510-521.

Del Rey, E., Racionero, M., 2010. Financing Schemes for Higher Education. European Journal of Political Economy 26(1), 104-113.

Eckwert, B., Zilcha, I., 2015. Student loans: When is risk sharing desirable? Forthcoming in International Journal of Economic Theory.

Eckwert, B., Zilcha, I, 2012, Private investment in higher education: Comparing alternative funding schemes. Economica (79), 76–96.

Friedman, M., 1962. Capitalism and Freedom. University of Chicago Press, Chicago.

Greenaway, D., Haynes, M., 2003. Funding higher education in the UK: The role of fees and loans. Economic Journal 113, F150–F166.

Hatsor, L., 2015, Higher-education funding: The value of information .Economic Letters 137, 230-233.

Landsberger, M., Meilijson, I., 1989. Demand for risky financial assets: A portfolio analysis. Journal of Economic Theory 50, 204-213.

Lleras, M. P., 2004. Investing in Human Capital. Cambridge University Press: Cambridge.

Milgrom, P. R., 1981. Good news and bad news: Representation theorems and applications. Bell Journal of Economics 12, 380-391.

Nerlove, M., 1975. Some problems in the use of income-contingent loans for the finance of higher education. Journal of Political Economy 83, 157-184.

Restuccia D., Urrutia, C., 2004. Intergenerational persistence of earnings: The role of early and college education. American Economic Review 94, 1354-1378.

Woodhall, M., 1988. Designing a student loan programme for a developing country: The relevance of international experience. Economics of Education Review 7, 153–61.